\documentclass[sigconf]{acmart}

\AtBeginDocument{%
  }

\setcopyright{acmcopyright}
\copyrightyear{2024}
\acmYear{2024}
\acmDOI{XXXXXXX.XXXXXXX}

\newtheorem{rem}{Remark}

\newcommand{\Real}{\mathbb R}
\newcommand{\eps}{\varepsilon}

\newcommand{\norm}[1]{\left\Vert#1\right\Vert}
\newcommand{\ra}{\rightarrow}
\newcommand{\set}[1]{\left\{#1\right\}}

\newcommand{\D}{\mathcal{D}}

\let\subset\subseteq

\usepackage{makecell}
\usepackage{enumitem}
\usepackage{listings}

\lstdefinestyle{mystyle}{
    language=Python,
    basicstyle=\ttfamily\small,
    commentstyle=\color{olive},
    keywordstyle=\color{blue},
    numberstyle=\tiny\color{gray},
    numbers=left,
    breaklines=true,
    showstringspaces=false,
}

\copyrightyear{2024}
\acmYear{2024}
\setcopyright{acmlicensed}\acmConference[HSCC '24]{27th ACM International Conference on Hybrid Systems: Computation and Control}{May 14--16, 2024}{Hong Kong, Hong Kong}
\acmBooktitle{27th ACM International Conference on Hybrid Systems: Computation and Control (HSCC '24), May 14--16, 2024, Hong Kong, Hong Kong}
\acmDOI{10.1145/3641513.3650134}
\acmISBN{979-8-4007-0522-9/24/05}

\begin{document}

\title{LyZNet: A Lightweight Python Tool for Learning and Verifying Neural Lyapunov Functions and Regions of Attraction}

\author{Jun Liu} 
\email{j.liu@uwaterloo.ca}
\affiliation{%
  \institution{University of Waterloo}
  \city{Waterloo}
  \country{Canada}
  }

\author{Yiming Meng}
\email{ymmeng@illinois.edu}
\affiliation{%
  \institution{University of Illinois}
  \city{Urbana}
  \country{United States}
}

\author{Maxwell Fitzsimmons}
\email{mfitzsimmons@uwaterloo.ca}
\affiliation{%
  \institution{University of Waterloo}
  \city{Waterloo}
  \country{Canada}
}

\author{Ruikun Zhou}
\email{ruikun.zhou@uwaterloo.ca}
\affiliation{%
  \institution{University of Waterloo}
  \city{Waterloo}
  \country{Canada}
}

\renewcommand{\shortauthors}{J. Liu, Y. Meng, M. Fitzsimmons, and R. Zhou}

\begin{abstract}
In this paper, we describe a lightweight Python framework that provides integrated learning and verification of neural Lyapunov functions for stability analysis. The proposed tool, named \textit{LyZNet}, learns neural \textbf{Ly}apunov functions using physics-informed neural \textbf{net}works (PINNs) to solve \textbf{Z}ubov's equation and verifies them using satisfiability modulo theories (SMT) solvers. What distinguishes this tool from others in the literature is its ability to provide verified regions of attraction close to the domain of attraction. This is achieved by encoding Zubov's partial differential equation (PDE) into the PINN approach. By embracing the non-convex nature of the underlying optimization problems, we demonstrate that in cases where convex optimization, such as semidefinite programming, fails to capture the domain of attraction, our neural network framework proves more successful. The tool also offers automatic decomposition of coupled nonlinear systems into a network of low-dimensional subsystems for compositional verification. We illustrate the tool's usage and effectiveness with several numerical examples, including both non-trivial low-dimensional nonlinear systems and high-dimensional systems.
\end{abstract}




\maketitle

\section{Introduction}

Stability analysis of nonlinear dynamical systems has been a focal point of research in control and dynamical systems. In many applications, characterizing the domain of attraction for an asymptotically stable equilibrium point is crucial. For example, in power systems, understanding the domain of attraction is essential for assessing whether the system can recover to a stable equilibrium after experiencing a fault. 

Since Lyapunov's landmark paper more than a hundred years ago \cite{lyapunov1992general}, Lyapunov functions have become a cornerstone of nonlinear stability analysis, providing an instrumental tool for performing such analyses and estimating the domain of attraction. Consequently, extensive research has been conducted on Lyapunov functions. One of the key technical challenges is the construction of Lyapunov functions. To address this challenge, both analytical \cite{haddad2008nonlinear,sepulchre2012constructive} and computational methods \cite{giesl2007construction, giesl2015review} have been investigated.

Among computational methods for Lyapunov functions, sums-of-squares (SOS) techniques have garnered widespread attention \cite{papachristodoulou2002construction,papachristodoulou2005tutorial,packard2010help,tan2008stability,topcu2008local,jones2021converse}. These methods facilitate not only local stability analysis but also provide estimates of regions of attraction \cite{topcu2008local,tan2008stability,packard2010help}. Leveraging semidefinite programming (SDP), one can extend the region of attraction by employing a specific ``shape function" within the estimated region. However, selecting such shape functions in a principled manner, beyond standard norm \cite{packard2010help} or quadratic functions \cite{khodadadi2014estimation}, remains elusive.

On the other hand, Zubov's theorem \cite{zubov1964methods} precisely characterizes the domain of attraction through a partial differential equation (PDE). This contrasts with commonly seen Lyapunov conditions, which manifest as partial differential inequalities. The use of an equation, rather than inequalities, enables precise characterization of the domain of attraction. The concept of maximal Lyapunov function \cite{vannelli1985maximal} is closely related to Zubov's method. The authors of \cite{vannelli1985maximal} have also provided a computational procedure for constructing maximal Lyapunov functions using rational functions.

\subsection{Related work}

The proposed tool solves Zubov's equation using physics-informed neural networks (PINNs) \cite{lagaris1998artificial,raissi2019physics} and verifies the neural solutions using satisfiability modulo theories (SMT) solvers. The tool currently relies on dReal \cite{gao2013dreal} for verification through interval analysis.

Computation of Lyapunov functions using Zubov's equation has been previously explored, for example, with radial basis functions \cite{giesl2007construction} and SOS techniques \cite{jones2021converse} (though the authors did not explicitly mention Zubov's equation). More recently, the authors of \cite{kang2021data} employed a data-driven approach to solve Zubov's equation using neural networks; however, the computed neural solutions were not verified, and Zubov's equation was not directly encoded in the loss function. The authors of \cite{liu2023towards} solved Zubov's equation using neural networks and verified them with dReal, but did not contrast their results with SOS techniques.

Many authors have recently investigated the use of neural networks for computing Lyapunov functions (see, e.g., \cite{chang2019neural,grune2021overcoming,gaby2022lyapunov,abate2020formal,gaby2022lyapunov,kang2021data}, and \cite{dawson2023safe} for a recent survey). In fact, such efforts date back to as early as the 1990s \cite{long1993feedback, prokhorov1994lyapunov}. Unlike SDP-based synthesis of SOS Lyapunov functions, neural network Lyapunov functions obtained by training are not guaranteed to be Lyapunov functions. Subsequent verification is required, e.g., using satisfiability modulo theories (SMT) solvers \cite{chang2019neural,ahmed2020automated}. The use of SMT solvers for searching and refining Lyapunov functions has been explored previously \cite{kapinski2014simulation}. Counterexample-guided search of Lyapunov functions using SMT solvers is investigated in \cite{ahmed2020automated} and the associated tool \cite{abate2021fossil}, which supports both Z3 \cite{de2008z3} and dReal \cite{gao2013dreal} as verifiers. Neural Lyapunov functions with SMT verification are explored in \cite{zhou2022neural} for systems with unknown dynamics. SMT verification is often time-consuming, especially when seeking a maximal Lyapunov function \cite{liu2023towards} or dealing with high-dimensional systems. Recent work has also focused on learning neural Lyapunov functions and verifying them through optimization-based techniques, e.g., \cite{chen2021learningHybrid,chen2021learningROA,dai2021lyapunov,dai2020counter}. Such techniques usually employ (leaky) ReLU networks and use mixed integer linear/quadratic programming (MILP/MIQP) for verification.

In contrast with existing approaches, the proposed tool offers the ability to compute regions of attraction (ROA) close to the domain of attraction and provides support for compositional verification to handle high-dimensional systems. We demonstrate its usage and effectiveness on both low-dimensional systems with non-trivial dynamics and high-dimensional systems to showcase compositional verification. The tool achieves state-of-the-art results on ROA approximations for nonlinear systems and their verification for high-dimensional systems.

\section{Problem formulation}

Consider a nonlinear system described by 
\begin{equation}
    \label{eq:sys}
    \dot{x}  = f(x),
\end{equation}
where \( f:\, \Real^n \ra \Real^n \) is continuously differentiable. Suppose that the origin is an asymptotically stable equilibrium for the system. We denote the unique solution to (\ref{eq:sys}) from the initial condition $x(0)=x_0$ by $\phi(t,x_0)$ for $t\in J$, where $J$ is the maximal interval of existence for $\phi$.  

The \textit{domain of attraction} of the origin for (\ref{eq:sys}) is defined as
\begin{equation}
    \D: = \set{x\in\Real^n:\,\lim_{t\ra \infty}\norm{\phi(t,x)} = 0}. 
\end{equation}
We know that $\D$ is an open and connected set. 

We call any forward invariant subset of $\D$ a \textit{region of attraction} (ROA). We are interested in computing regions of attraction, as they not only provide a set of initial conditions with guaranteed convergence to the equilibrium point, but also ensure constraints and safety through forward invariance. 

Lyapunov functions provide an indispensable tool for stability analysis and ROA estimates. A \textit{Lyapunov function} for asymptotic stability of the origin is a continuously differentiable function satisfying $V(0)=0$, $V(x)>0$ for all $x\neq 0$, and the derivative of V along solutions of (\ref{eq:sys}) (sometimes called its Lie derivative) satisfies 
$$
\dot V(x):= \frac{dV}{dx}\cdot f(x) < 0,\quad \forall x\neq 0.
$$
Given any such function and any positive constant $c>0$, the sublevel set $\set{x\in\Real^n:\, V(x)\le c}$ is a region of attraction, provided that the set is contained in $\D$. 

The proposed tool is designed to accomplish the following task: given an asymptotically stable equilibrium point, it computes neural Lyapunov functions that can certify regions of attraction with provable guarantees.

What sets our tool and approach apart from existing ones in the literature is its ability to verify regions of attraction close to the domain of attraction by incorporating Zubov's equation into the training of neural Lyapunov functions.

\section{Tool overview and usage}

\subsection{Tool overview}

The tool consists of a set of Python modules built around the machine learning framework PyTorch \cite{paszke2019pytorch} and the numerical satisfiability modulo theories (SMT) solver dReal \cite{gao2013dreal}. It also relies on the NumPy library \cite{harris2020array} for numerical computation, SciPy library \cite{virtanen2020scipy} for scientific computing tasks, and the SymPy library \cite{meurer2017sympy} for symbolic calculation. 

We describe the main modules as follows:
\begin{itemize}[left=0cm]

\item \texttt{dynamical\_systems.py} defines the \texttt{DynamicalSystems} class and associated methods such as linearization, computation of quadratic Lyapunov function by solving Lyapunov's  equation, and providing interfaces for numerical and symbolic function valuations related to the vector field. 

\item \texttt{local\_verifier.py} provides functions for verifying local stability using quadratic Lyapunov functions, with interval analysis from the SMT solver dReal. The key component is using the mean value theorem to bound the remainder of the linear approximation. This enables \textit{exact} verification of local asymptotic stability, despite dReal's numerical conservativeness. This sets it apart from existing methods that often ignore verification around a neighbourhood of the origin.

\item \texttt{quadratic\_verifier.py} performs reachability analysis using a quadratic Lyapunov function. The target is the local ROA obtained above. The set guaranteed to reach this target is formulated as a sublevel set of the quadratic Lyapunov function.

\item \texttt{generate\_data.py} generates data for training neural Lyapunov functions. It involves solving system (\ref{eq:sys}), augmented with an additional state to compute value for the solution to Zubov's equation, which we accomplish using SciPy and ``embarrassing parallelization'' with the joblib library \cite{joblib}.

\item \texttt{neural\_learner.py} learns a neural Lyapunov function by solving Zubov's PDE using physics-informed neural networks (PINN) \cite{lagaris1998artificial,raissi2019physics}. We allow customization of neural network structure as well as selection of different loss function modes. 

\item \texttt{neural\_verifier.py} verifies the learned neural Lyapunov function indeed satisfies the Lyapunov condition required for reach a target invariant set within the domain of attraction. The target can be set as the region of attraction already verified by local analysis or quadratic Lyapunov analysis above. Our current implementation uses dReal as the verifier. 

\item \texttt{network\_verifier.py} includes functions for compositional verification of local stability using quadratic and neural Lyapunov functions. The underlying theory involves vector Lyapunov functions, differential inequalities, and comparison techniques for nonlinear systems.

\item \texttt{sos\_verifier.py}, \texttt{plot.py}, and \texttt{utils.py} provide support for comparisons with other Lyapunov functions, such as sums-of-squares (SOS) Lyapunov functions, visualization of verified level sets, and other routines used in learning and verification.

\end{itemize}

\subsection{Tool usage}
\label{sec:usage}

Using the tool is straightforward. As shown in the code snippet below, we can define a dynamical system with SymPy variables and symbolic expressions for the vector field. We can also specify the domain for learning and training, as well as name the system for file management and result storage.
\begin{lstlisting}[style=mystyle, frame=single]
import sympy 
import lyznet

# Define dynamics
mu = 1.0
x1, x2 = sympy.symbols('x1 x2')
f_vdp = [-x2, x1 - mu * (1 - x1**2) * x2]
domain_vdp = [[-2.5, 2.5], [-3.5, 3.5]]
sys_name = "Van_der_Pol"
vdp_system = lyznet.DynamicalSystem(f_vdp, domain_vdp, sys_name)
\end{lstlisting}

The code above defines the reversed Van der Pol equation with parameter $\mu=1.0$ on the domain $X=[-2.5, 2.5]\times [-3.5,3.5]$. 

The following lines of code can be used to call relevant functions for verifying local stability using quadratic Lyapunov function, as well as learning and verifying neural Lyapunov functions. 

\begin{lstlisting}[style=mystyle, frame=single]
# Call the local stability verifier
c1_P = lyznet.local_stability_verifier(vdp_system)
# Call the quadratic verifier
c2_P = lyznet.quadratic_reach_verifier(vdp_system, c1_P)

# Generate data 
data = lyznet.generate_data(vdp_system, n_samples=3000)

# Call the neural lyapunov learner
net, model_path = lyznet.neural_learner(vdp_system, data=data, lr=0.001, layer=2, width=30, num_colloc_pts=300000, max_epoch=20, loss_mode="Zubov")

# Call the neural lyapunov verifier
c1_V, c2_V = lyznet.neural_verifier(vdp_system, net, c2_P)
\end{lstlisting}

Most functions and parameters are self-explanatory. We remark that the argument \texttt{loss\_mode} in \texttt{neural\_learner} currently can be set to three different modes: \texttt{"Zubov"}, \texttt{"Data"}, and \texttt{"Lyapunov"}. The latter two take a purely data-driven approach for solving Zubov's equation or encode a Lyapunov inequality instead of Zubov's equation, respectively.

The code above computes and verifies four sublevel sets:
\begin{equation}\label{eq:levels}
\begin{aligned}
    \mathcal{P}_1 & = \set{x\in X:\, V_P(x)\le {c1}\_P}, \\
    \mathcal{P}_2 & = \set{x\in X:\, V_P(x)\le {c2}\_P}, \\
    \mathcal{V}_1 & = \set{x\in X:\, V_N(x)\le {c1}\_V}, \\
    \mathcal{V}_2 & = \set{x\in X:\, V_N(x)\le {c2}\_V},
    \\
\end{aligned}
\end{equation}
satisfying $\mathcal{P}_1\subset \mathcal{P}_2$, $\mathcal{V}_1\subset\mathcal{P}_2$, and $\mathcal{V}_1\subset \mathcal{V}_2$, where $V_P$ is a quadratic Lyapunov function of the form $V_P(x)=x^TPx$ and $V_N(x)$ is a learned neural network Lyapunov function. The set $\mathcal{P}_1$ is a local region of attraction verified by linearization. The set $\mathcal{P}_2$ represents the set of initial conditions from which solutions of (\ref{eq:sys}) are verified to reach $\mathcal{P}_1$ using a quadratic Lyapunov function.  The sets $\mathcal{V}_1$ and $\mathcal{V}_2$ are verified using a neural Lyapunov function, where $\mathcal{V}_1$ represents a target invariant set contained in $\mathcal{P}_2$\footnote{This could be set to $\mathcal{P}_1$ if one wishes to skip the call of \texttt{quadratic\_reach\_verifier} for any reason, or any known region of attraction verified by other means.}, and $\mathcal{V}_2$ is an invariant set from which solutions of (\ref{eq:sys}) are guaranteed to reach $\mathcal{V}_1$. All sets are also verified to be contained in the interior of $X$ to ensure set invariance. Put together, $\mathcal{V}_2$ is a verified ROA for (\ref{eq:sys}). 

\section{Methodology}
\label{sec:method}

\subsection{Local stability analysis using quadratic Lyapunov functions}\label{sec:local_stability}

Local stability analysis using quadratic Lyapunov functions is fairly standard. Suppose that $x=0$ is an exponentially stable equilibrium point for (\ref{eq:sys}); i.e., \( f(0)=0 \) and the Jacobian matrix \( A=Df(0)=\frac{df}{dx}\big\vert_{x=0} \) is a Hurwitz matrix, i.e., all eigenvalues of \( A \) have negative real parts. Rewrite (\ref{eq:sys}) as 
\begin{equation}
    \label{eq:nonlinear}
    \dot x = Ax + g(x),
\end{equation}
where $g(x)=f(x)-Ax$ satisfies $\lim_{x\ra 0}\frac{\norm{g(x)}}{\norm{x}}=0$. Given any symmetric positive definite real matrix $Q\in\Real^{n\times n}$, there exists a unique solution $P$ to the Lyapunov equation 
\begin{equation}
    \label{eq:lyap_linear}
    PA +A^T P = -Q.
\end{equation}
Let $V_P(x)=x^TPx$. Then 
\begin{align*}
\dot{V}_P(x) & = x^T(PA+A^TP)x + 2x^TPg(x)\\
&= - x^TQx + 2x^TPg(x) \\
& \le - \lambda_{\min}(Q)\norm{x}^2 + 2x^TPg(x) \\
& = -\eps \norm{x}^2  + (2x^TPg(x) - r\norm{x}^2), 
\end{align*}
where we set $r=\lambda_{\min}(Q)-\eps>0$ for some sufficiently small $\eps>0$ and $\lambda_{\min}(Q)$ is the minimum eigenvalue of $Q$. 
The goal of \texttt{local\_stability\_verifier} is to determine the largest number $c1\_P$ such that 
\begin{equation}\label{eq:c1_P}
x\in \mathcal{P}_1 \Longrightarrow 2x^TPg(x)\le r\norm{x}^2, 
\end{equation}
where $\mathcal{P}_1$ is as defind in (\ref{eq:levels}).

While (\ref{eq:c1_P}) can be easily formulated as a satisfiability condition in dReal \cite{gao2013dreal}, due to conservative use of interval analysis in dReal to account for numerical errors, verification of inequalities such as $2x^TPg(x)\le r\norm{x}^2$ will return a counterexample close to the origin. To overcome this issue, we look at a higher-order approximation of $Pg(x)$. By the mean value theorem, we have
$$
    P g(x) = P g(x) - P g(0) = \int_0^1 P\cdot Dg(tx)dt\cdot x,
$$
where $Dg$ is the Jacobian of $g$ given by $Dg= Df - A$, 
which implies
$$
 2x^TP g(x) \le 2\sup_{0\le t\le 1}\norm{P \cdot Dg(tx)}\norm{x}^2.
$$
As a result, to verify (\ref{eq:c1_P}), we just need to verify 
\begin{align}
x\in \mathcal{P}_1 &\Longrightarrow 2\sup_{0\le t\le 1}\norm{P\cdot Dg(tx)}\le r.\label{eq:cr2}
\end{align}
If $\mathcal{P}_1$ is star-shaped with respect to the origin\footnote{A set $S$ is star-shaped with respect to the point $x_0$ if for all $y\in S$ the line segment joining $y$ and $x_0$ is a subset of $S$.}, then (\ref{eq:cr2}) is equivalent to 
\begin{align}
x\in \mathcal{P}_1 &\Longrightarrow 2\norm{P\cdot Dg(x)}\le r.\label{eq:cr3}
\end{align}
Since $Dg(0)=0$ and $Dg$ is continuous, one can always choose ${c1}\_P>0$ sufficiently small such that (\ref{eq:cr3}) can be verified. Furthermore, in rare situations, if (\ref{eq:cr3}) can be verified for $X=\Real^n$, then global exponential stability of the origin is proved for (\ref{eq:sys}). 

\begin{rem}
From (\ref{eq:cr3}), one can further use easily computable norms, such as the Frobenius norm, to over-approximate the matrix 2-norm $\norm{P\cdot Dg(tx)}$. Furthermore, by the implication ${x^TPx\le {c1}\_P} \Longrightarrow |x_i|\le \frac{{c1}\_P}{\lambda_{\min}(P)}$ for all $i=1,\cdots,n$, and the use of a compositionally verifiable upper bound of the matrix 2-norm, such as the Frobenius norm, one can verify (\ref{eq:cr3}) in a compositional fashion. This idea was implemented in \texttt{compositional\_local\_stability\_verifier}, which can work significantly more efficiently for high-dimensional systems. Of course, this decomposition of an ellipsoid set to a hyperrectangular set is inherently conservative. Less conservative results for compositional verification can be achieved with the help of vector Lyapunov functions \cite{liu2024compositionally}, as implemented in \texttt{network\_verifier}. 
\end{rem}

\subsection{Learning neural Lyapunov functions via Zubuv's equation for ROA verification}\label{sec:zubov}

\subsubsection{Zubov's theorem and maximal Lyapunov function}

Zubov's PDE \cite{zubov1964methods} takes the form
\begin{equation}
    \label{eq:zubov}
    \dot W(x):= \nabla W(x) \cdot f(x) = - \Psi(x) (1-W(x)), 
\end{equation}
where $W$ is a positive definite function to be solved and $\Psi$ is also a positive definite function that can be chosen. For instance, $\Psi$ can be chosen as $\Psi(x)=\alpha (1+W(x)) \norm{x}^2$, where $\alpha>0$ is a parameter. Zubov's theorem states that if there exists a function $W$ satisfying the above PDE on a given domain $D$ containing the origin with the following two additional conditions:  (1) $0< W(x)< 1$ on $D$ except $W(0)=0$; (2) $W(x)\ra 1$ as $x\ra \partial D$, then the domain $D=\mathcal D$, the domain of attraction for the origin. Suppose that the function $W$ is also trivially extended to the domain of interest $X$ containing $D$ with $W(x)=1$ for $x\in X\setminus D$. We have $\D=\set{x\in X:\,W(x)< 1} $. In other words, the domain of attraction is characterized by the sublevel-1 set of the solution to Zubov's PDE. 

Zubov's theorem is closely related to the notion of maximal Lyapunov function discussed in \cite{vannelli1985maximal}. A maximal Lyapunov function on a domain $D$ containing the origin satisfies the property 
\begin{equation}
    \label{eq:lyap}
    \dot V(x):= \nabla V(x) \cdot f(x) = - \omega(x),
\end{equation}
for all $x\in D$, where $\omega$ is a positive definite function. Additionally, $V(x)\ra \infty$ as $x\ra \partial D$. 

A solution $W$ to Zubov's equation can be related to a maximum Lyapunov $V$ by any strictly increasingly function $\beta:\,[0,\infty)\ra \Real$ that satisfies $\beta(0)=0$ and $\beta(s)\ra 1$ as $s\ra\infty$. Indeed, given a maximal Lyapunov function $V$, we can simply choose 
\begin{equation}
    \label{eq:beta}
W(x) = \beta(V(x)). 
\end{equation}
There are obvious choices of such a function $\beta$. For examples $\beta(s) =1 - \exp(-\alpha s)$ for $s\ge 0$ or $\beta(s)=\tanh(\alpha s)$ for $s\ge 0$, where $\alpha>0$ is a parameter. In our implementation, we choose $\beta(s)=\tanh(\alpha s)$. It can be easily verified that $W$ defined this way satisfies (\ref{eq:zubov}) with 
\begin{equation}\label{eq:tanh}
\Psi(x)=-\alpha (1+W(x))\omega(x). 
\end{equation}

Under suitable assumptions \cite{vannelli1985maximal}, a solution to (\ref{eq:lyap}) can be constructed as 
\begin{equation}\label{eq:int_data}
V(x) = \int_0^\infty \omega(\phi(t,x))dt.     
\end{equation}
For example, if the origin is an exponentially stable equilibrium point and $\omega$ is locally Lipschitz, then this construction is valid and gives a maximal Lyapunov function on $\mathcal{D}$. One easy choice of $\omega$ is $\omega(x)=\norm{x}^2$. 

\subsubsection{Physics-informed neural solution to Zubov's PDE}

Put together, (\ref{eq:zubov}), (\ref{eq:tanh}), and (\ref{eq:int_data}) allow us to learn a neural Lyapunov function via physics-informed neural networks \cite{lagaris1998artificial,raissi2019physics} for solving PDEs.  

In a nutshell, a physics-informed neural solution to the PDE (\ref{eq:zubov}) is a neural network function that minimizes the residual for satisfying (\ref{eq:zubov}). Additional conditions on this neural network function can be enforced (or rather encouraged) through additional loss terms.

More specifically, let $W_N(x;\theta)$ be a neural approximation to a solution of (\ref{eq:zubov}). The training loss consists of 
\begin{equation}
    \label{eq:loss}
    \mathcal{L}(\theta) = \mathcal{L}_{\text{residual}}(\theta)  + \mathcal{L}_{\text{boundary}}(\theta) + \mathcal{L}_{\text{data}}(\theta), 
\end{equation}
where $\mathcal{L}_{\text{residual}}$ is the residual error of the PDE given by
\begin{equation}
    \label{eq:L_r}
    \frac{1}{N}\sum_{i=1}^N(\nabla_{x} W_N(x_i;\theta) f(x_i) + \Psi(x_i)(1-W_N(x_i;\theta)))^2,  
\end{equation}
evaluated over a set of collocation points $\set{x_i}_{i=1}^{N}$ chosen over a domain $X$. For instance, we can choose $\Psi$ as in (\ref{eq:tanh}). The loss $\mathcal{L}_{\text{boundary}}$ captures boundary conditions. There are different ways boundary conditions can be added. The simplest one is that $W(0)=0$ and $W(x)=1$ for $x\not\in \mathcal{D}$. Although optional, we can also encourage the following inequality \cite{grune2021computing,liu2023towards} near the origin:
\begin{equation} \label{eq:c1Wc2}
\beta(c_1\norm{x}^2) \le W(x) \le \beta(c_2 \norm{x}^2), 
\end{equation}
where $\beta(s)=\tanh(\alpha s)$ (corresponding to the choice of $\beta$ above). This inequality is always satisfiable for some positive $c_1$ and $c_2$ when the origin is exponentially stable. Finally, the term $\mathcal{L}_{\text{data}}(\theta)$ captures data loss, which we define as
\begin{equation}
    \label{eq:L_d}
    \mathcal{L}_{\text{data}}(\theta) = \frac{1}{N_d}\sum_{i=1}^{N_d}(W_N(y_i;\theta)-\hat W(y_i))^2,
\end{equation}
where $\{\hat W(y_i)\}_{i=1}^{N_d}$ is a set of data points, which can be obtained by forward integration of (\ref{eq:sys}) to obtain $W$ from (\ref{eq:int_data}) and (\ref{eq:beta}) \cite{kang2021data}.  

\subsubsection{Verification of regions of attraction}

We briefly outline how local stability analysis via linearization and reachability analysis via a Lyapunov function can be combined to provide a ROA estimate that is close to the domain of attraction. 

Let $V_P$ and $W_N$ be two generic functions. Suppose that $V_P$ and $c>0$ gives a local region of attraction $\set{x\in X:\,V_P(x)\le c}$. We can verify the following inequalities using SMT solvers: 
\begin{align}
(c_1\le {W}_N(x) \le c_2) \wedge (x\in X) &\Longrightarrow \dot{W}_N(x)  \le -\eps,\label{eq:dW}\\
({W}_N(x)\le c_1) \wedge (x\in X)  &\Longrightarrow V_P(x)\le c,\label{eq:WP}
\end{align}
where $\eps>0$, $c_2>c_1>0$. The first inequality ensures that solutions of (\ref{eq:sys}) starting in $\set{x\in X:\,W_N\le c_2}$ always reaches the target set $\set{x\in X:\,W_N\le c_1}$, as long as the solution does not leave $X$. The second inequality ensures that the target set $\set{x\in X:\,W_N\le c_1}$ is contained in a local region of attraction $\set{x\in X:\,V_P(x)\le c}$. By definition, it follows that $\set{x\in X:\,W_N\le c_2}$ provides an under-estimate of the true domain of attraction $\mathcal{D}$. An easy condition to ensure that solutions cannot leave $X$ before reaching the target is that $\set{x\in X:\,W_N\le c_2}$ does not intersect with the boundary of $X$, which can also be easily verified by an SMT solver. 

Now specific to the tool, we can use the above technique to first compute a region of attraction via linearization as in Section \ref{sec:local_stability} and obtain $c1\_P$. We can then use the reachability conditions above to obtain an enlarged region of attraction, first using a quadratic Lyapunov function to verify $c2\_P$, and then using a neural Lyapunov function to verify $c1\_V$ and $c2\_V$, as shown in (\ref{eq:levels}). If the neural Lyapunov function $V_N$ is obtained by solving Zubov's PDE (\ref{eq:zubov}), then the set $\mathcal{V}_2:=\set{x\in X:\, V_N(x)\le c2\_V}$ provides a region of attraction close to the true domain of attraction. 

\begin{rem}
While there are numerous studies on computing neural Lyapunov functions \cite{abate2020formal,abate2021fossil,ahmed2020automated,grune2021computing,chang2019neural,zhou2022neural,gaby2022lyapunov}, the tools currently available cannot produce verified region of attraction estimates that are close to the state-of-the-art SOS approaches. In particular, we tested the recent tool Fossil 2.0 \cite{abate2021fossil,edwards2023fossil} on Example \ref{ex:vdp} in the following section and found that the region of attraction verified by Fossil is worse than that verified by a quadratic Lyapunov function (dashed red curve in Figure \ref{fig:vdp1}) using LyZNet. On the other hand, in the next section, we demonstrate LyZNet's capability in outperforming state-of-the-art SOS approaches by providing less conservative verified region-of-attraction estimates. This is because LyZNet aims to solve Zubov's equation, which precisely characterizes the exact domain of attraction.
\end{rem}

\section{Examples}\label{sec:examples}

In this section, we demonstrate the usage of the tool with several numerical examples. All experiments are conducted on a 2020 MacBook Pro with a 2 GHz Quad-Core Intel Core i5 and without any GPU. The purpose of this section is not to conduct extensive experiments, but to demonstrate the usage and effectiveness of the proposed tool. The equations describing the examples can be found in the Appendix. We note that the main purpose here is to demonstrate the capability of the tool. Readers are referred to \cite{liu2023physics,liu2024compositionally} for more detailed discussions of the numerical results and, especially, compositional verification techniques involved \cite{liu2024compositionally}. The repository of the tool includes code for running these examples (\url{https://git.uwaterloo.ca/hybrid-systems-lab/lyznet}). 

\begin{figure}[h]
  \centering
  \includegraphics[width=\linewidth]{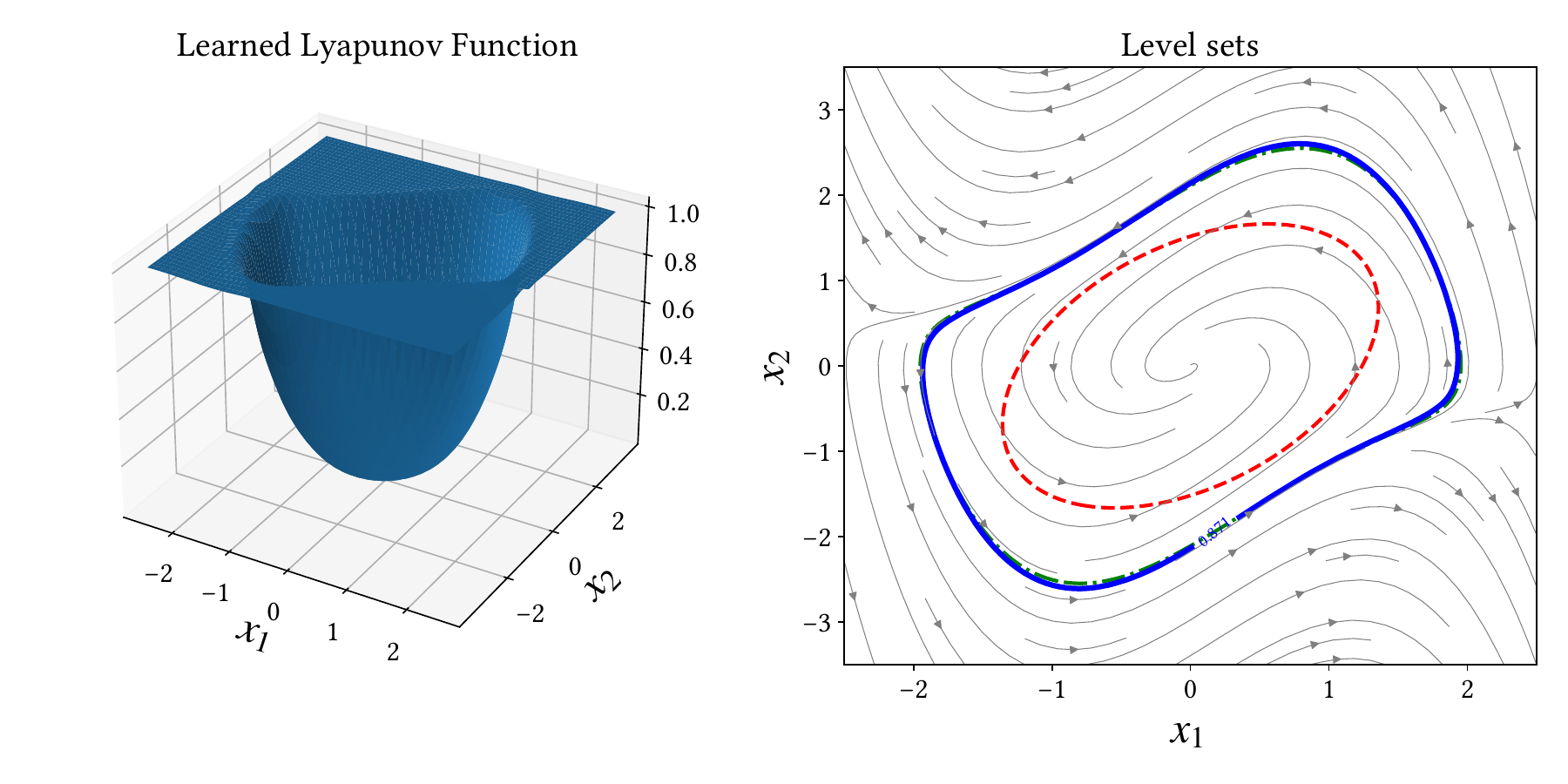}
  \caption{Verified neural Lyapunov function for Van der Pol equation with $\mu=1.0$ (Example \ref{ex:vdp}). Dashed red: quadratic Lyapunov function; dot-dashed green: SOS Lyapunov; solid blue: neural Lyapunov.}
  \label{fig:vdp1}
\end{figure}

\begin{example}[Van der Pol equation]\label{ex:vdp}
    Consider the reversed Van der Pol equation with different values of $\mu$. For $\mu = 1.0$ and $X = [-2.5, 2.5] \times [-3.5, 3.5]$, similar to the code snippet shown in Section \ref{sec:usage}, we train two neural networks of different sizes (depth and width). Figure \ref{fig:vdp1} depicts the largest verifiable sublevel set, along with the learned neural Lyapunov function. It can be seen that the domain of attraction is quite comparable and slightly better than that provided by a sums-of-squares (SOS) Lyapunov function with a polynomial degree of 6, obtained using a standard ``interior expanding" algorithm \cite{packard2010help}. 
\end{example}

\begin{figure}[h]
  \centering
  \includegraphics[width=\linewidth]{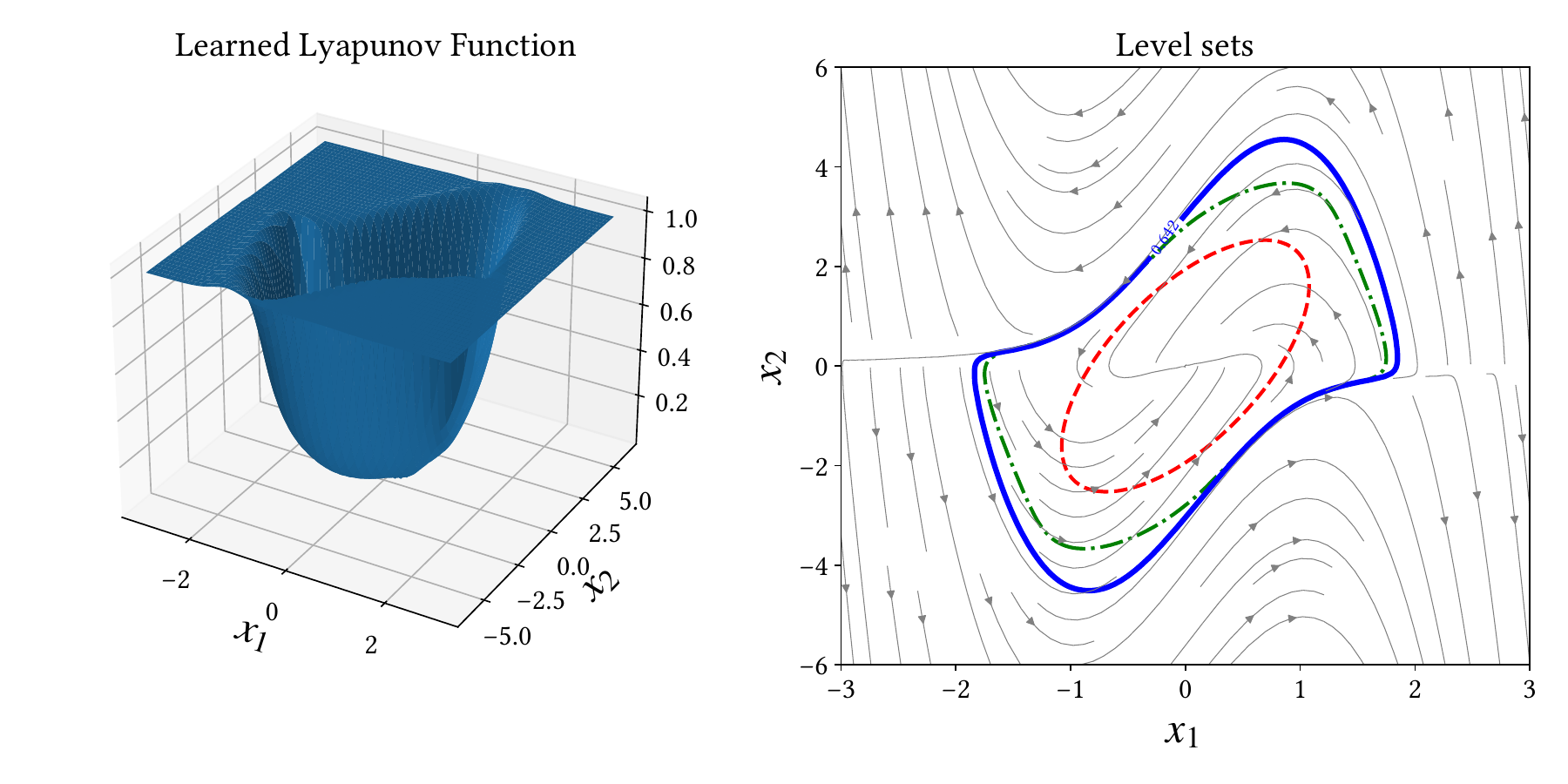}
  \caption{Verified neural Lyapunov function for Van der Pol equation with $\mu=3.0$ (Example \ref{ex:vdp}). Dashed red: quadratic Lyapunov function; dot-dashed green: SOS Lyapunov; solid blue: neural Lyapunov. The learned neural Lyapunov function outperforms a sixth degree SOS Lyapunov function.}
  \label{fig:vdp3}
\end{figure}

As the stiffness of the equation increases, we observe further improved advantages of neural Lyapunov approach over SOS Lyapunov functions. With $\mu=3.0$ and domain $X=[-3.0, 3.0]\times [-6.0, 6.0]$, the comparison of neural Lyapunov function with SOS Lyapunov function is shown in Figire \ref{fig:vdp3}.  

\begin{table}
  \caption{Parameters and verification results for Van der Pol equation (Example \ref{ex:vdp})}
  \label{tab:vdp}
  \begin{tabular}{cccccc}
    \toprule
 $\mu$  & Layer & Width & c2\_V & Volume (\%) 
 & SOS volume (\%)\\ 
    \midrule
    1.0 & 2 & 30 & 0.871 & \textbf{94.78\%} &  94.10\% \\
    \midrule
    3.0 & 2 & 30 & 0.642  & \textbf{85.21\%} &  70.93\%\\
    \bottomrule
  \end{tabular}
\end{table}

\begin{figure}[h]
  \centering
  \includegraphics[width=\linewidth]{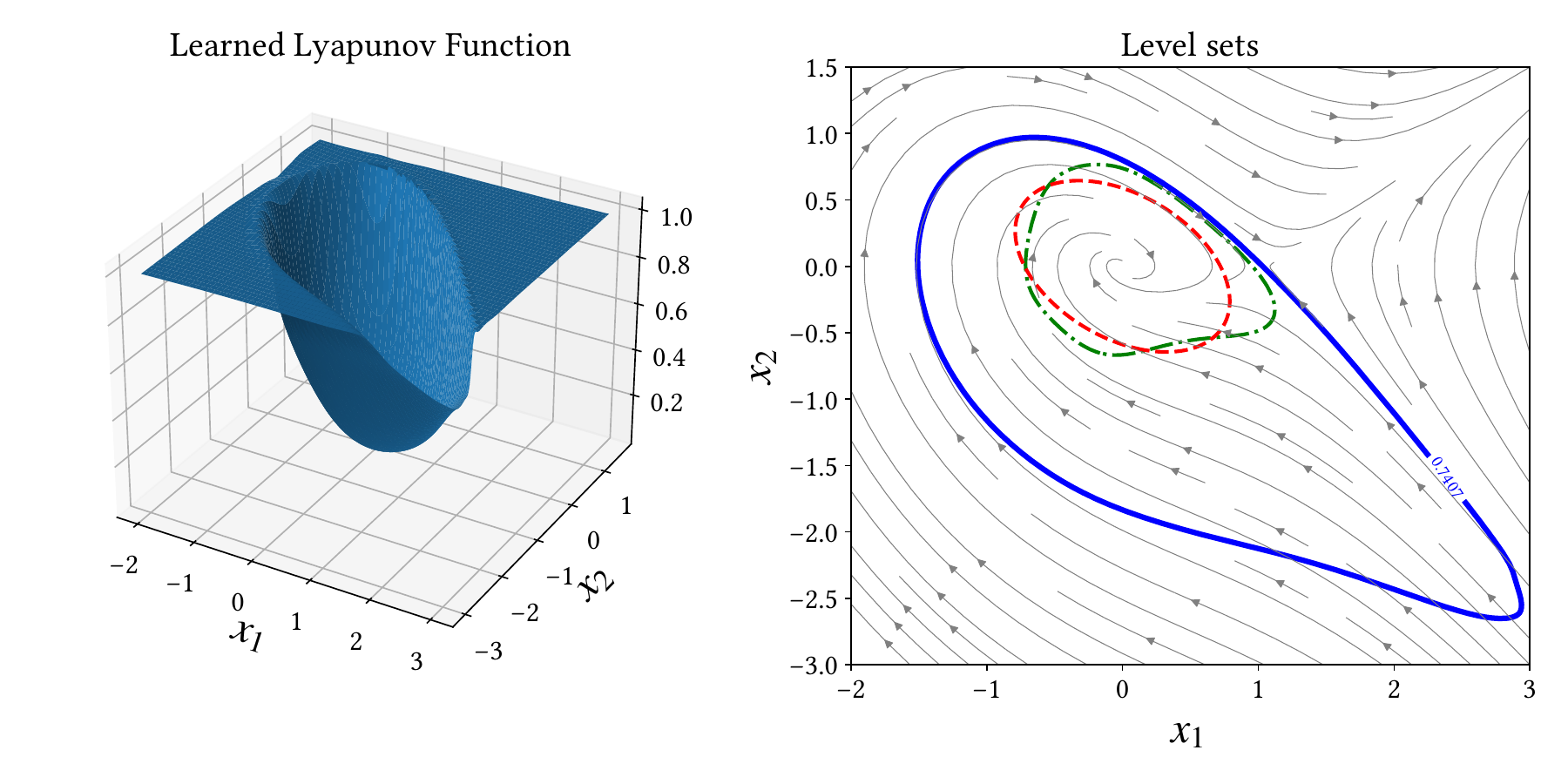}
  \caption{Verified neural Lyapunov function for a two-machine power system (Example \ref{ex:power}). Dashed red: quadratic Lyapunov function; dot-dashed green: SOS Lyapunov; solid blue: neural Lyapunov. The learned neural Lyapunov function significantly outperforms a sixth degree SOS Lyapunov function.} 
  \label{fig:power}
\end{figure}

\begin{example}[Two-machine power system]\label{ex:power} Consider a two-machine power system  \cite{vannelli1985maximal} which has an asymptotically stable equilibrium point at the origin and an unstable equilibrium point at $(\pi/3,0)$. Figure \ref{fig:power} shows that a neural network with two hidden layers and 30 neurons in each layer provides a region-of-attraction estimate significantly better than that from a sixth-degree polynomial SOS Lyapunov function, computed with a Taylor expansion of the system model. The example shows that neural Lyapunov functions perform better than SOS Lyapunov functions when the nonlinearity is non-polynomial. We also compared with the rational Lyapunov function presented in \cite{vannelli1985maximal}, but the ROA estimate is worse than that from the SOS Lyapunov function, and we were not able to formally verify the sublevel set of the rational Lyapunov function reported in \cite{vannelli1985maximal} with dReal \cite{gao2013dreal}. Improving the degree of polynomial in the SOS approach does not seem to improve the result either. 
\end{example}

\begin{table}
  \caption{Parameters and verification results for a two-machine power system (Example \ref{ex:power})}
  \label{tab:power}
  \begin{tabular}{ccccc}
    \toprule
  Layer & Width & c2\_V & Volume (\%) 
 & SOS volume (\%)\\ 
    \midrule
 2 & 30 & 0.740 & \textbf{82.34\%} &  18.53\% \\
    \bottomrule
  \end{tabular}
\end{table}

\begin{example}[Inverted pendulum with linear control] In the literature, there have been numerous references to this example as a benchmark for comparing techniques for stabilization with provable ROA estimates. Interestingly, our \texttt{local\_stability\_verifier} verifies global stability of the origin within 1 millisecond using dReal \cite{gao2013dreal}. 
\end{example}

\begin{rem}\label{rem:time}
Computational time for training and verification largely depends on computer hardware. On the somewhat outdated laptop used for experiments, training the neural network as shown in Tables \ref{tab:vdp} and \ref{tab:power} takes about 500 seconds. This is with 20 epochs and 300,000 collocation points using a minibatch size of 32. Verification, including all bisection subroutines, can range from 200 to 2,000 seconds. We report the computational times in the Appendix. 
\end{rem}

\begin{example}[10-dimensional system]\label{ex:10d}
The example was used by the authors of \cite{grune2021computing} and \cite{gaby2022lyapunov} to illustrate the training of neural network Lyapunov functions for stability analysis, where the trained Lyapunov functions were not verified. Here, we use the example to illustrate the compositional verification functionalities of LyZNet. 

For the compositional approach, we examine two decompositions. The first splits the system into five linear subsystems, each with two variables. The nonlinear terms are treated as interconnections. The second has ten linear subsystems of the form $\dot{x}_i = -x_i$, each with one variable, and considers the remaining terms as interconnections. LyZNet automates these decompositions. 

We tried verification over different domains with different approaches. The results are summarized in Table \ref{tab:10d}. It can be seen that for a smaller domain $X=[-1,1]^{10}$, all verifiers work well and find the largest possible sublevel sets contained in the domain. It is worth noting that only the decomposition ``$10\times 1$ dim'' gives a verified ROA equal to the entire domain (barring the conservativeness of the numerical SMT verifier dReal~\cite{gao2013dreal}), because the other two approaches either produce the largest Euclidean ball or the largest Cartesian product of Euclidean balls contained in the domain.

As the domain gets larger, it becomes more challenging for the monolith approach. For $X=[-10,10]^{10}$, the monolith approach takes more than 12 hrs without returning a result, while all compositional approaches succeeded. Notably, the ``$10\times 1$ dim'' decomposition provides the largest possible hyper-rectangular invariance set within the domain; indeed, it can be verified this set is $[-5,5]^{10}$. 

We also trained neural network functions and observe that, if we restrict the structural complexity of the neural network, we are able to verify regions of attraction using neural Lyapunov functions. However, the region of attraction is not as good as the ones obtained using quadratic Lyapunov functions, as those ones are already optimal for this example. 

\begin{table}
  \caption{Verification results for a 10-dimensional system using quadratic Lyapunov functions (Example \ref{ex:power})}
  \label{tab:10d}
  \begin{tabular}{ccccc}
    \toprule
  Domain & Approach & Verifier & Levels & Time (sec)\\ 
    \midrule
 $[-1,1]^{10}$ & monolith & local & 0.49999 & 0.19\\
  & \makecell{compositional\\ ($5\times 2\text{ dim}$)} & local & 0.49999 & 0.12\\
 & \makecell{compositional\\ ($10\times 1\text{ dim}$)} & local & 0.49999 & 0.29\\
 \midrule 
 $[-4,4]^{10}$ & monolith & local & 7.99999 & 4.30 \\ 
  & \makecell{compositional\\ ($5\times 2\text{ dim}$)} & local & 7.99999 & 0.12\\
  & \makecell{compositional\\ ($10\times 1\text{ dim}$)} & local & 3.1210 & 4.45\\
  & \makecell{compositional\\ ($10\times 1\text{ dim}$)} & quadratic & 7.99999 & 1.54\\
  \midrule
 $[-10,10]^{10}$ & monolith & local & -- & \makecell{time out\\ 
 (>43200 sec)} \\ 
  & \makecell{compositional\\ ($5\times 2\text{ dim}$)} & local & 12.4938 & 2.07\\
  & \makecell{compositional\\ ($10\times 1\text{ dim}$)} & local & 3.1188 & 4.57 \\
  & \makecell{compositional\\ ($10\times 1\text{ dim}$)} & quadratic & 12.4969 & 23.40 \\
    \bottomrule
  \end{tabular}
\end{table}

\end{example}

\section{Conclusions and future work}

We presented a lightweight Python framework for learning and verifying neural Lyapunov functions. We demonstrated that by solving Zubov's PDE using neural networks, the verified region of attraction can indeed approach the boundary of the domain of attraction, outperforming sums-of-squares Lyapunov functions obtained using semidefinite programming. To cope with learning and verifying Lyapunov functions for high-dimensional systems, we have built support for compositional verification and demonstrated its effectiveness over a monolithic approach. While not presented in this paper, the compositional approach has been demonstrated to be effective for learning and verifying neural Lyapunov functions \cite{liu2024compositionally}. 

There are numerous ways the tool can be expanded. Ongoing research focuses on compositional training and verification of neural Lyapunov functions \cite{liu2024compositionally}. Future work could include supporting verification engines other than dReal, such as Z3 \cite{de2008z3}, and leveraging the growing literature on neural network verification tools. Extending support to other types of dynamical systems like delay differential equations and stochastic dynamics is also of interest. Exploring other stability and boundedness notions, such as global stability and ultimate boundedness, could be valuable. For compositional verification, designing neural networks that automatically discover compositional structures in high-dimensional systems is a promising direction, as opposed to the current manual assignment. Integrating convex optimization and semidefinite programming for suitable problems is another avenue. Data-driven computation of verifiable Lyapunov functions using Zubov's equation is another interesting direction \cite{meng2023learning,meng2024zubov}. Finally, the tool has the potential to handle controls, making the simultaneous training of controllers and Lyapunov/value functions a natural next step. Initial results have been reported in \cite{meng2024physics} and to appear in \cite{liu2024lyznet_control}.  

\begin{acks}
This research was supported in part by the Natural Sciences and Engineering Research Council of Canada and the Canada Research Chairs program. The development of the tool was also enabled in part by support provided by the Digital Research Alliance of Canada (alliance.ca).
\end{acks}

\bibliographystyle{ACM-Reference-Format}
\bibliography{hscc24}

\section{Appendix}

This appendix presents the dynamic equations for the numerical examples discussed in Section \ref{sec:examples}. Additionally, we report the computational times required for the training and verification processes in Examples \ref{ex:vdp} and \ref{ex:power}.

\subsection{Van der Pol equation}

The reversed Van der Pol equation is given by
    \begin{equation}
\begin{aligned}
        \dot{x}_1 & = -x_2, \\
        \dot{x}_2 & = x_1 - \mu (1 - x_1^2) x_2,
\end{aligned}
    \end{equation}
where $\mu > 0$ is a parameter that affects the stiffness of the equation.

\subsection{Two-machine power system}

Consider the two-machine power system \cite{vannelli1985maximal} modelled by 
    \begin{equation}
\begin{aligned}
        \dot{x}_1 & = x_2, \\
        \dot{x}_2 & = -0.5x_2 - (\sin(x_1 +\delta)-\sin(\delta)),
\end{aligned}
    \end{equation}
    where $\delta = \frac{\pi}{3}$. 
    
\subsection{Inverted pendulum with linear control}

Consider an inverted pendulum 
    \begin{equation}
\begin{aligned}
        \dot{x}_1 & = x_2, \\
        \dot{x}_2 & = \sin(x_1) - x_2 - (k_1 x_1 + k_2 x_2),
\end{aligned}
    \end{equation}
where the linear gains are given by $k_1 = [4.4142, 2.3163]$. 

\subsection{Synthetic 10-dimensional system} 

Consider the 10-dimensional nonlinear system from \cite{grune2021computing}:
\begin{align*}
 \dot x_1 =    &-x_1 + 0.5 x_2 - 0.1 x_9^2, \\
 \dot x_2 =   &-0.5 x_1 - x_2, \\
 \dot x_3 =    &-x_3 + 0.5 x_4 - 0.1 x_1^2, \\
 \dot x_4 =    &-0.5 x_3 - x_4, \\
 \dot x_5 =  &-x_5 + 0.5 x_6 + 0.1 x_7^2, \\
 \dot x_6 =   &-0.5 x_5 - x_6, \\
 \dot x_7 =   &-x_7 + 0.5 x_8, \\
 \dot x_8 =   &-0.5 x_7 - x_8, \\
 \dot x_9 =   &-x_9 + 0.5 x_{10}, \\
 \dot x_{10} =    &-0.5 x_9 - x_{10} + 0.1 x_2^2.
\end{align*}

\subsection{Computational times}

\begin{table}[h!]
  \caption{Training and verification times for Examples \ref{ex:vdp} (Van der Pol) and  \ref{ex:power} (two-machine power)}
  \label{tab:time}
  \begin{tabular}{ccc}
    \toprule
Model &  Training (sec) & Verification (sec) \\ 
    \midrule
Van der Pol ($\mu=1$) & 494 & 973\\
Van der Pol ($\mu=3$) & 502 & 442\\
two-machine power & 473 & 3375  \\
\bottomrule
\end{tabular}
\end{table}

We report the computational times required for the training and verification processes in Examples \ref{ex:vdp} and \ref{ex:power} on a 2020 MacBook Pro with a 2 GHz Quad-Core Intel Core i5, without any GPU. For training, we randomly chose 300,000 collocation points in the domain and trained for 20 epochs. The sizes of the neural networks were as reported in Tables \ref{tab:vdp} and \ref{tab:power}. Verification in dReal can be easily parallelized by adjusting the config parameter \texttt{number\_of\_jobs}, which may correspond to the number of cores/threads on the computer. For the laptop used in these experiments, there are four cores, allowing for eight threads via hyperthreading. As discussed in Remark \ref{rem:time}, computational time is largely dependent on computer hardware. The computational times provided in Table \ref{tab:time} serve as a reference point. We expect that training time can be significantly improved by using GPUs, and verification time can be further reduced by utilizing additional cores. 

\end{document}